\documentclass[%
reprint,
superscriptaddress,
showpacs,preprintnumbers,
verbatim,
amsmath,amssymb,
aps,
pre,
]{revtex4-1}

\usepackage{color}
\usepackage{graphicx}% Include figure files
\usepackage{dcolumn}% Align table columns on decimal point
\usepackage{bm}% bold math
\begin {document}
%section {title}
%\preprint{APS/123-QED}

\title{%
  A piecewise linear model of self-organized hierarchy formation}
% Force line breaks with \\

%% Time-averaged mean-square-displacement tensor: a novel method to elucidate
%% fluctuating diffusivity

\author{Tomoshige Miyaguchi}
\email{tmiyaguchi@naruto-u.ac.jp}
\affiliation{%
  Department of Mathematics,
  Naruto University of Education, Tokushima 772-8502, Japan}
\author{Takamasa Miki}
\affiliation{%
  Department of Mathematics,
  Naruto University of Education, Tokushima 772-8502, Japan}
\author{Ryota Hamada}
\affiliation{%
  Department of Mathematics,
  Naruto University of Education, Tokushima 772-8502, Japan}

%% \author{Takuma Akimoto}
%\email{akimoto@keio.jp}
%% \affiliation{%
%% Department of Mechanical Engineering, Keio University, Yokohama, 223-8522, Japan
%% }%

%%   \author{Eiji Yamamoto}
%%   \affiliation{%
%%   Department of Mechanical Engineering, Keio University, Yokohama, 223-8522, Japan
%% }%


\date{\today}% It is always \today, today,

\begin{abstract}
  The Bonabeau model of self-organized hierarchy formation is studied by using a
  piecewise linear approximation to the sigmoid function. Simulations of the
  piecewise-linear agent model show that there exist two-level and three-level
  hierarchical solutions, and that each agent exhibits a transition from
  non-ergodic to ergodic behaviors. Furthermore, by using a mean-field
  approximation to the agent model, it is analytically shown that there are
  asymmetric two-level solutions, even though the model equation is symmetric
  (asymmetry is introduced only through the initial conditions), and that
  linearly stable and unstable three-level solutions coexist. It is also shown
  that some of these solutions emerge through supercritical-pitchfork-like
  bifurcations in invariant subspaces. Existence and stability of the linear
  hierarchy solution in the mean-field model are also elucidated.
\end{abstract}

%\pacs{05.45.Ac, 05.40.Fb, 87.15.Vv}% PACS, the Physics and Astronomy
\maketitle

%\tableofcontents

\section {Introduction}
%subsection {general introduction}
%subsubsection {establish importance of this research topic}
Hierarchy formation has been intensively studied in a wide range of animal
species: insects \cite{oliveira90}, fish \cite{goessmann00, chase02,
  grosenick07}, birds \cite{lindquist09}, and mammals \cite{wittig03} including
even humans \cite{savin80, garandeau14}.
%subsubsection {provide general background information}
It has been considered that not only differences in the prior attributes of
individuals such as weight and aggressiveness but also social interactions
between individuals are important in the hierarchy formation \cite{chase02,
  castellano09}. In fact, it is known that an individual who won an earlier
contest has a higher probability of winning later contests than an individual
who lost the earlier contest (winner-loser effects) \cite{hsu06}. Positive
feedback generated through such effects might enhance the formation of
hierarchies in animal groups.

To elucidate such a feedback mechanism in hierarchy formations, a mathematical
model is proposed by Bonabeau et al \cite{bonabeau95}. The Bonabeau model
consists of $N$ agents, and each agent $i\,(i=1,\dots,N)$ is characterized by a
variable $F_i(t)$, where $t$ is time. $F_i(t)$, which is called strength or
fitness in the literature, is referred to as a dominance score (DS) in this
paper \cite{lindquist09}. After a contest between two agents $i$ and $j$,
$F_i(t)$ increases if the agent $i$ wins and decreases if $i$ loses [the same
rule is applied to $F_j(t)$]. A greater value of $F_i(t)$ means a higher
probability to win a contest. In addition, the agents are assumed to perform
random walks on a two-dimensional square lattice $L \times L$, and a contest
occurs when two agents meet; thus, the density of the agents $\rho = N/L^2$ is a
parameter, which controls the frequency of contests. In addition to these
pairwise interactions, $F_i(t)$ is assumed to show a relaxation according to a
differential equation $dF_i(t) / dt = -\mu \tanh \bigl(F_i(t)\bigr)$.

It is found that, as the density $\rho$ increases, the Bonabeau model shows a
transition from an egalitarian state in which all $F_i(t)$ are equal to a
hierarchical state in which $F_i(t) \neq F_j(t)$ for some $i\neq j$. The
Bonabeau model is one of the basic models of the hierarchy formation, and it is
compared with experimental observations of hierarchies in animal groups
\cite{bonabeau99, lindquist09}. Hierarchical structures can be well described by
the Bonabeau model \cite{bonabeau99}, but some discrepancies are also reported
\cite{lindquist09}.

%subsubsection {describe general problem or current research focus of the field}

Since the Bonabeau model is a simple model, many modified versions have been
proposed to make it more realistic. In Refs.~\cite{stauffer03a, stauffer03b},
another feedback mechanism and an asymmetric rule are incorporated into the
dynamics of $F_i(t)$. This generalized model (a Stauffer version) is analyzed in
Ref.~\cite{lacasa06}, and it was found that the egalitarian solution is always
stable, while a two-level stable solution (a hierarchical solution) appears at a
critical parameter value through a saddle-node bifurcation. In addition, a model
with a simpler relaxation dynamics $dF_i(t) / dt = - \mu F_i(t)$ is also
analyzed in Ref.~\cite{lacasa06}, and it was found that a similar transition
occurs but in this case the bifurcation is supercritical-pitchfork
type. Recently, an asymmetric model is intensively studied in
Ref.~\cite{posfai18}; In this asymmetric model, each agent has an intrinsic
parameter called a talent, which can be considered as a prior attribute of that
agent. Moreover, two modified models are proposed in Refs.\cite{odagaki06,
  tsujiguchi07, okubo07}: a timid-society model and a challenging-society
model. In the timid-society model, an agent can choose a vacant site when it
moves and thereby it can avoid a contest; in the challenging-society model, the
agent chooses the strongest neighbor as an opponent.

%subsection {literature review to solve the general problem}
%subsubsection {provide a brief overview of key research projects}

In contrast to these modifications trying to incorporate realistic features,
there are also works intending to simplify the Bonabeau model \cite{bennaim05,
  bennaim06, bennaim07}. In these studies, the DSs of agents are assumed to
attain only integer values, and the DS of the winner increases by one and that
of the loser does not change. The dynamics can be described by a partial
differential equation in a continuum limit. This model also shows a transition
from the egalitarian solution to a hierarchical solution.

%subsubsection {describe a gap in the research}

In spite of this diversity of models of the hierarchy formation, understanding
of the original Bonabeau model is still limited. For example, in the Bonabeau
model with relaxation dynamics $dF_i(t) / dt = - \mu F_i(t)$, it is found that
the egalitarian solution is stable at low densities (at small values of $\rho$).
This egalitarian solution becomes unstable at $\rho = \rho_c$, and two-level
stable solutions appear through a supercritical-pitchfork bifurcation
\cite{lacasa06}. But, it seems impossible to rigorously derive the stable range
of this two-level solution (some approximation is necessary).  This difficulty
stems mainly from nonlinearity of the sigmoid function employed in the Bonabeau
model (See Sec.~\ref{s.models}).

%subsection {describe the present paper}
%subsubsection {describe the present paper}

In this paper, we propose another simplified version of the Bonabeau model by
introducing a piecewise linear function in place of the sigmoid function.
%subsubsection {describe the methodology used in the present paper}
%paragraph {theory}
%paragraph {model}
Piecewise linear approximations are often used in the studies of nonlinear
dynamical systems. In fact, even for systems in which rigorous approaches are
difficult, more detailed analysis is possible for piecewise linear versions
\cite{devaney84, tasaki02, miyaguchi06, miyaguchi07b}.
%subsubsection {announce the findings}
Here, we derive the stable ranges of two-level and three-level solutions for the
piecewise-linear model. Moreover, we found that asymmetric two-level solutions
exist even though the system is symmetric (asymmetry is introduced only through
the initial conditions). It is also shown that various stable and unstable
solutions coexist.

%subsubsection {organization of the paper}

This paper is organized as follows. In Sec.~\ref{s.models}, we define two
piecewise-linear models of the hierarchy formation: an agent model and a
mean-field model. In Sec.~\ref{s.linear-stability}, linear stability analysis
for steady solutions (i.e., fixed points \cite{strogatz94}) of the mean-field
model is presented. In Sec.~\ref{s.ergodicity-breaking}, a transition from
ergodic to non-ergodic behaviors in the agent model is numerically studied.
Finally, Sec.~\ref{s.summary} is devoted to a discussion, in which we suggest
possible generalizations of the agent model.

\section {Models}\label{s.models}
%subsection {intro}

In this section, we introduce two models of the self-organized hierarchy
formation. The first model is referred to as an agent model, and the second as a
mean-field model. It is shown that the mean-field model is a good approximation
of the agent model in a weak interaction limit.

\subsection {Agent model}\label{s.agent-model}
%subsubsection {intro}

Let us suppose that there are $N$ agents, and each agent $i \,\,(i=1,\dots,N)$
is characterized by a real number $F_i(t)$, which is referred to as the DS at
time $t$ \cite{bonabeau95}. $F_i(t)$ is a measure of strength or fitness of the
agent $i$ \cite{castellano09, lacasa06}, and changes through interactions with
other agents; Hereafter, the interaction between two agents is referred to as a
contest. Firstly, we define the dynamics just at the contest; Secondly, we
define inter-contest dynamics by using a Poisson process.

%subsubsection {fig1}
\begin{figure}[t]
  \centerline{\includegraphics[width=5.5cm]{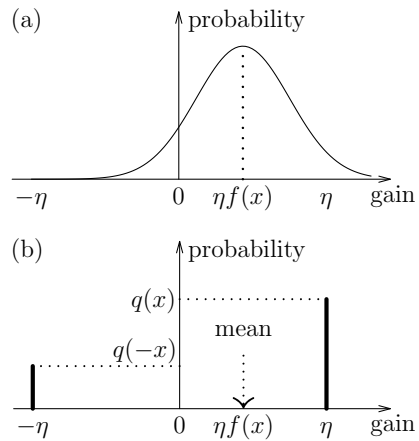}}
  %% \centerline{\includegraphics[width=5.5cm]{bullying.gain.eps}}
  %% \centerline{\includegraphics[width=5.5cm]{bullying.gain.bonabeau.eps}}
  \caption{\label{f.gain}Probabilities of the gain $F_i(t_n^+) - F_i(t_n^-)$ (a)
    for the model defined in Eq.~(\ref{e.agent_model.Fi}), and (b) for the
    Bonabeau model. Here, $x$ stands for the difference of the DSs of two
    contestants, e.g., $x = F_i(t^-_n) - F_j(t^-_n)$. Then, $q(x) \,[q(-x)]$ is
    the probability that the agent $i$ wins (loses) the contest with $j$. The
    mean gain $\mu$ is expressed as $\mu = \eta [q(x) - q(-x)] = \eta f(x)$
    [Eq.~(\ref{e.mean-gain})].}
\end{figure}

\subsubsection {Contest dynamics}

Let us define the dynamics of $F_i(t)$ at the contests. At random time $t=t_n$,
two agents $i$ and $j$ contest with each other, where $i$ and $j$ are
randomly chosen from the $N$ agents. In this contest, the values of $F_i(t)$ and
$F_j(t)$ change as
\begin{align}
  \label{e.agent_model.Fi}
  F_i(t_n^+) &= F_i(t_n^-) + \eta f\bigl(F_i(t^-_n) - F_j(t^-_n)\bigr) + \xi_i(t_n),
  \\[0.1cm]
  \label{e.agent_model.Fj}
  F_j(t_n^+) &= F_j(t_n^-) + \eta f\bigl(F_j(t^-_n) - F_i(t^-_n)\bigr) + \xi_j(t_n),
\end{align}
where $t_n^-$ and $t_n^+$ are the times just before and after the contest,
respectively. A gain from winning or losing the contest is defined by the
difference of the DSs before and after the contest, $F_i(t_n^+) - F_i(t_n^-)$,
which is equivalent to the sum of the second and the third terms in the right
side of Eq.~(\ref{e.agent_model.Fi}). Therefore, the parameter $\eta$ controls
the amount of the gain, thereby characterizing the impact of the contest
result. Moreover, $\xi_i(t)$ is a random variable following the normal
distribution with mean $0$ and variance $\sigma^2$, and satisfies an
independence property
$\left\langle \xi_i(t_n) \xi_j(t_m)\right\rangle = \delta_{ij} \delta_{nm}
\sigma^2$. The gain is thus a random variable, and its probability density is
illustrated in Fig.~\ref{f.gain}(a).

In Eqs.~(\ref{e.agent_model.Fi}) and (\ref{e.agent_model.Fj}), $f(x)$ is a
non-linear function similar to the sigmoid function. In this paper, we assume it
has the following piecewise linear form:
\begin{align}
  \label{e.def.f(x)}
  f(x)=
  \begin{cases}
    -1              & (x\leq -2F_0),
    \\[0.1cm]
    \frac{x}{2F_0} & (-2F_0 \leq x \leq 2F_0),
    \\[0.1cm]
    1               & (x \geq 2F_0),
  \end{cases}
\end{align}
where $F_0$ characterizes the scale of $F_i(t)$, and it can be removed by
rescaling [See Appendix \ref{s.rescale}]. Note also that $x$ stands for the DS
difference of two contestants, e.g., $x = F_i(t^-_n) - F_j(t^-_n)$ [See
Eq.~(\ref{e.agent_model.Fi})]. This function $f(x)$ is a piecewise-linear
approximation to the function
\begin{equation}
  \label{e.f(x).bonabeau}
  f_{\mathrm{b}}(x) = \frac {1}{1+e^{- x/F_0}} - \frac {1}{1+e^{x/F_0}}.
\end{equation}
This function $f_{\mathrm{b}}(x)$ is employed in the original Bonabeau model
\cite{bonabeau95}. Due to the nonlinearity in $f_{\mathrm{b}}(x)$, theoretical
analysis of the Bonabeau model is difficult except for a few simple steady
solutions. For the piecewise linear approximation given by
Eq.~(\ref{e.def.f(x)}), however, more detailed analysis of hierarchical
solutions is possible due to its simplicity.

The first and the second terms on the right side of Eq.~(\ref{e.f(x).bonabeau})
have a simple probabilistic interpretation.  Let us define a function $q(x)$ as
$q(x) := [f_b(x)+1]/2$, then Eq.~(\ref{e.f(x).bonabeau}) can be expressed as
$f_b(x) = q(x) - q(-x)$.  If $x$ is given by $x = F_i(t^-_n) - F_j(t^-_n)$, the
first term $q(x)$ is the winning probability of $i$ against $j$, and the second
term $q(-x)$ is the losing probability of $i$ against $j$. A similar
interpretation is possible also for our model [Eq.~(\ref{e.def.f(x)})]; If we
rewrite $f(x)$ as $f(x) = q(x) - q(-x)$ with $q(x) = [f(x) + 1]/2$, then $q(x)$
[$q(-x)$] is the probability of winning (losing). Thus, $\eta f(x)$ in the right
side of Eq.~(\ref{e.agent_model.Fi}) is the mean gain of the agent $i$ through
the contest with $j$.

%subsubsection {explanation of fig1}

Apart from this difference in $f(x)$ and $f_b(x)$, Eqs.~(\ref{e.agent_model.Fi})
and (\ref{e.agent_model.Fj}) are still slightly different from the Bonabeau
model, for which the dynamics is given by
\begin{equation}
  F_i(t_n^+) = F_i(t_n^-) \pm \eta
\end{equation}
with the plus sign if the agent $i$ wins and the minus sign if it loses (a
similar equation holds for the opponent). The probabilities of winning and
losing are given by $q(x)$ and $q(-x)$ defined above. Thus, the gain of the
contest is a random variable following a dichotomous distribution as shown in
Fig.~\ref{f.gain}(b).

The present model shown in Fig.~\ref{f.gain}(a) can be considered as a
coarse-grained version of the Bonabeau model. This is because a sum of several
gains, each following the dichotomous distribution in Fig.~\ref{f.gain}(b),
should follow a continuous distribution similar to the one in
Fig.~\ref{f.gain}(a) by virtue of the central limit theorem
\cite{feller71}. Therefore, our model might well be plausible for some species
for which the same pair of individuals contest in succession \cite{lindquist09}.

More precisely, if we assume that the same pair contests $T$ times in succession
in the Bonabeau model, the noise terms in Eqs.~(\ref{e.agent_model.Fi}) and
(\ref{e.agent_model.Fj}) can be considered as small. In fact, the mean and the
variance of the sum of the dichotomous gains approximately become
\begin{align}
  \mu      &\approx T \eta \left[q(x) - q(-x)\right],\\[0.1cm]
  \sigma^2 &\approx 4 T \eta^2 q(x)q(-x).
\end{align}
Let us rescale $\eta$ by replacing it with $\eta/T$, we obtain
\begin{align}
  \label{e.mean-gain}
  \mu      &\approx\eta \left[q(x) - q(-x)\right], \\[0.1cm]
  \label{e.mean-variance}
  \sigma^2 &\approx \frac {4 \eta^2}{T} q(x)q(-x).
\end{align}
This is the situation shown in Fig.~\ref{f.gain}(a). From
Eqs.~(\ref{e.mean-gain}) and (\ref{e.mean-variance}), it is found that, if the
timescale $T$ is large, the standard deviation $\sigma$ can be considered as
small compared with the mean value $\mu$. Therefore, in the following, we study
the simplest case $\sigma^2 = 0$ and neglect the noise terms $\xi_i(t)$ and
$\xi_j(t)$ in Eqs.~(\ref{e.agent_model.Fi}) and (\ref{e.agent_model.Fj}). Note
also that this noiseless model can be considered as a simplification of the
Bonabeau model in that the random dichotomous gains $\pm \eta$ in the Bonabeau
model are replaced by its mean value $\mu$ given in Eq.~(\ref{e.mean-gain}) [See
Fig.~\ref{f.gain}(b)].

%% Note however that $\sigma^2$ should then depend on $\triangle$, but we
%% neglect this dependence.

%% In this sense, the impact parameter $\eta$ and the variance $\sigma^2$ should
%% depend on the lag time $\tau_{n+1}$ (see below), but we assume these
%% parameters are constant for brevity.

%subsubsection {fig2}

\begin{figure}[t]
  \centerline{\includegraphics[width=8.7cm]{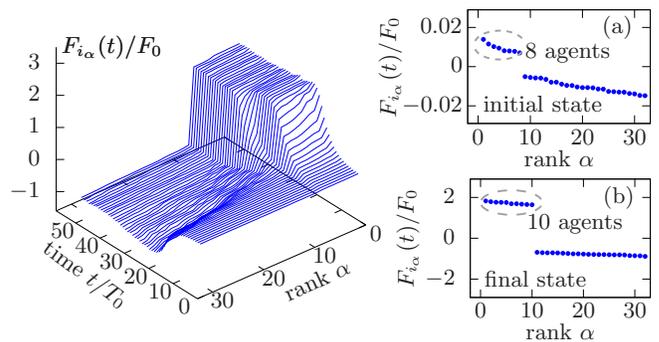}}
  %% \begin{minipage}[bt]{5.cm}
  %%   \centerline{\includegraphics[width=5cm]{m8-3d.f.eps}}
  %% \end{minipage}
  %% \begin{minipage}[bt]{3.5cm}
  %%   \centerline{\includegraphics[width=3.5cm]{m8-initial.f.eps}}
  %%   \centerline{\includegraphics[width=3.5cm]{m8-end.f.eps}}
  %% \end{minipage}
  \caption{\label{f.agent.two-level} (Left) Two-level hierarchy formation in the
    agent model with $N=32$. Time evolution of the DS profile
    $F_{i_{\alpha}}(t)$ is displayed as a function of the rank $\alpha$ and time
    $t$. Here, $i_{\alpha}(t)$ is the agent index of which rank is $\alpha$ at
    time $t$. The parameters $\eta$ and
    $\gamma$ are set as $\eta=10^{-3}F_0$ and $\gamma \eta =1.3 \rho_c$ with
    $\rho_c$ given by Eq.~(\ref{e.rho_c.general}).
    %% At every step of the numerical integration, the
    %% dominance score $F_i(t)$ is sorted so that $F_i(t)$ is decresing with
    %% increasing $i$.
    (Right) The initial and final DS profiles are shown in (a) and (b),
    respectively.}
\end{figure}

\subsubsection {Inter-contest dynamics}

In addition to the dynamics just at the contests, we should define the
inter-contest dynamics. We assume that the contests occur at random times
$t=t_1, \cdots, t_n, \cdots$ (we set $t_0=0$ for convenience). In the Bonabeau
model, the agents are assumed to perform random walks, and the times $t_n$ are
determined by random encounters of the agents \cite{bonabeau95}. However, the
random walk model introduces non-trivial correlations in the sequence of the
intervals $\tau_n := t_{n} - t_{n-1}\, (n=1,2,\dots )$.

Here, however, we assume that these intervals $\tau_n$ are mutually independent
random variables, and follow the exponential distribution:
\begin{equation}
  \label{e.exp.dist}
  w(\tau) =  \gamma_a e^{- \gamma_a \tau},
\end{equation}
where $\gamma_a$ is the interaction rate and its inverse $1/\gamma_a$ is the
mean of $\tau$. Thus, the inter-contest dynamics is the Poisson process
\cite{feller71}, and simplifies the model dynamics thanks to the independence of
the intervals $\tau_n$. In the original Bonabeau model, the contest is
considered as a diffusion-limited reaction, while the Poisson process might
arise from a reaction-limited random walk.

Note that $\gamma_a dt$ is the mean number of contests in the time interval
$dt$. Then, the mean number of contests in which the agent $i$ involves is
$\gamma_a dt \times 2/N$. Therefore, let us define $\gamma := 2\gamma_a/N$,
which is an interaction rate for a single agent.  In Appendix \ref{s.rescale},
we show that $\eta$ and $\gamma_a$ (or $\gamma$) completely characterize the
agent model.

Relaxation of the dominance relationship is observed in experiments of animal
groups. For example, in Ref.~\cite{chase02}, a group of fish is assembled to
form a hierarchy, then each individual in the group is separated for long time,
and finally they are assembled to form a hierarchy again. This second hierarchy
is often different from the first, and thus it is considered that individual
fish forgets the earlier dominance relationship.

Therefore, in the meantime of the contests in our model, $F_i(t)$ is assumed to
decay. As a relaxation dynamics, we employ the following differential equation:
\begin{equation}
  \label{e.relaxation}
  \frac {dF_i(t)}{dt} = - \frac {F_i(t)}{T_0}.
\end{equation}
Here, $T_0 > 0$ is a characteristic time scale of the relaxation, and it can be
removed by rescaling (See Appendix \ref{s.rescale}).

%subsubsection {fig3}
\begin{figure}[t]
  \centerline{\includegraphics[width=8.7cm]{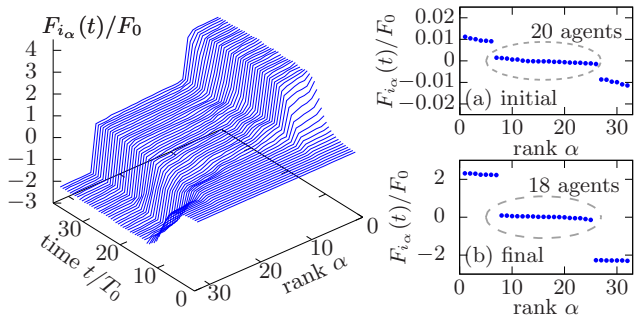}}
  %% \begin{minipage}[bt]{5.cm}
  %%   \centerline{\includegraphics[width=5cm]{m10-3d.f.eps}}
  %% \end{minipage}
  %% \begin{minipage}[bt]{3.5cm}
  %%   \centerline{\includegraphics[width=3.5cm]{m10-initial.f.eps}}
  %%   \centerline{\includegraphics[width=3.5cm]{m10-end.f.eps}}
  %% \end{minipage}
  \caption{\label{f.agent.three-level} (Left) Three-level hierarchy formation in
    the agent model with $N=32$. Time evolution of the DS profile
    $F_{i_{\alpha}}(t)$ is displayed as a function of the rank $\alpha$ and time
    $t$. The parameters $\eta$ and $\gamma$ are set as $\eta=10^{-3}F_0$ and
    $\gamma \eta=1.5 \rho_c$ with $\rho_c$ given by
    Eq.~(\ref{e.rho_c.general}). (Right) The initial and final DS profiles are
    shown in (a) and (b), respectively.}
\end{figure}

%subsubsection {fig: explanation}

In Figs.~\ref{f.agent.two-level} and \ref{f.agent.three-level}, results of
numerical simulations for the agent model are presented. In these simulations,
we neglect the noise terms $\xi_i(t)$ (i.e., we set $\sigma^2 = 0$). The initial
condition $F_i(0)$ is weakly stratified into two and three groups as shown in
Figs.~\ref{f.agent.two-level}(a) and \ref{f.agent.three-level}(a), respectively.
At long times, the DS profile $F_i(t)$ converges to stratified profiles slightly
different from the initial profiles (but there are some fluctuations at the
final states due to stochastic dynamics, i.e., the random sampling of the
contestants, and the random intervals $\tau_n$). As shown in
Fig.~\ref{f.agent.two-level}(b), the final state is an asymmetric two-level
profile, whereas in Fig.~\ref{f.agent.three-level}(b), the final state is a
symmetric three-level profile. In addition, even if the parameters are the same,
there are several final profiles depending only on the initial conditions and
realizations of the stochastic dynamics. Therefore, it is conjectured that
several stable profiles coexist at the same parameter values.

\subsection {Mean-filed model}\label{s.mean-field-model}

To analyze the stable profiles in the agent model, the mean-field model has been
employed in previous works \cite{bonabeau95}. In contrast to the agent model,
which is a stochastic model, the mean field model is deterministic and thus
described by ordinary differential equations. Here, let us apply the mean-field
approximation to the agent model introduced in the previous subsection.

If $1/\gamma \ll T_0$, there are many contests between the agent $i$ and the
other agents in the time scale $T_0$. In addition, if $\eta \ll F_0$
\footnote{More precisely, $\eta \gamma \delta t\ll F_0$ should be satisfied. If
  we can choose $\delta t$ satisfying this condition with
  $1/\gamma \ll \delta t \ll T_0$, the approximation in
  Eq.~(\ref{e.mean-field-approx}) is valid.}, then $F_i(t)$ does not change
greatly (compared with $F_0$) in each contest. Under these assumptions, changes
of $F_i(t)$, denoted as $\delta F_i(t)$, due to contests in the interval
$(t, t+ \delta t)$ ($1/\gamma \ll \delta t \ll T_0$) can be approximated as
\begin{equation}
  \label{e.mean-field-approx}
  \eta \sum_{k=1}^{\gamma \delta t} f\bigl(F_i(t) - F_{j_k}(t)\bigr)
  \approx
  \gamma \delta t \frac {\eta}{N'}
  \sum_{\begin{subarray}{c}j=1\\j\neq i \end{subarray}}^{N}
  f\bigl(F_i(t) - F_{j}(t)\bigr), 
\end{equation}
where $N'$ is defined as $N' := N-1$ and $j_k$ is the index of the $k$-th
contestant of $i$ in the interval $\delta t$.

By incorporating the relaxation term [Eq.~(\ref{e.relaxation})], the dynamics of
$F_i(t)$ can be described by the ordinary differential equations
\begin{align}
  \frac {dF_i(t)}{dt}
  %% &\approx
  %% \frac {\xi_{ij}(t)}{\tau_{0}} - \gamma F_i(t)
  %% \\[0.1cm]
  %% &\approx
  %% - \frac {F_i(t)}{T_0}
  %% +
  %% \frac {\rho \eta}{(N-1)}\sum_{\begin{subarray}{c}j=1\\j\neq i\end{subarray}}^{N}
  %% \xi_{ij}(t)
  %% \notag\\[0.1cm]
  \label{e.def.mean-field-model}
  &\approx
  - \frac {F_{i}(t)}{T_0}
  +
  \frac {\gamma \eta}{N'}
  \sum_{\begin{subarray}{c}j=1\\j\neq i\end{subarray}}^{N}f\bigl(F_i(t) - F_j(t)\bigr).
\end{align}
This equation (\ref{e.def.mean-field-model}) is the same form as the Bonabeau's
mean-field model \cite{bonabeau95}, in which the function $f(x)$ is given by
Eq.~(\ref{e.f(x).bonabeau}). Here, however, we employ the piecewise linear
function given in Eq.~(\ref{e.def.f(x)}).
%
%% In the following, we set $T_0=1$ and $F_0 = 1$ for brevity, but this does not
%% lose generality of the original mean-field model in
%% Eq.~(\ref{e.def.mean-field-model}) (See Appendix \ref{s.rescale}).

\section {Linear stability analysis of mean-field model}\label{s.linear-stability}
%subsection {intro}

In this section, we study steady solutions of the mean-field model with
$T_0 = 1$:
\begin{align}
  \label{e.mean-field-model.non-dimension}
  \frac {dF_i(t)}{dt}
  =
  - F_{i}(t)
  +
  \frac {\rho}{N'}
  \sum_{\begin{subarray}{c}j=1\\j\neq i\end{subarray}}^{N}
  f\bigl(F_i(t) - F_j(t)\bigr),
\end{align}
where $\rho \geq 0$ is defined as $\rho = \gamma\eta$. Moreover, $f(x)$ in
Eq.~(\ref{e.mean-field-model.non-dimension}) is assumed to be given by
Eq.~(\ref{e.def.f(x)}) with $F_0 = 1$ [i.e., Eq.~(\ref{e.def.f(x).nondimension})
in Appendix \ref{s.rescale}]. In Appendix \ref{s.rescale}, it is shown that such
simplifications do not lead to loss of generality, and that $\rho$ is the only
parameter of the mean-field model. In the figures, however, we give the units
explicitly.

It can be shown that the total DS defined by $S(t)=\sum_{i=1}^{N}F_i(t)$ follows
the equation $dS/dt = -S$, and thus $S(t)$ decays to zero as $t\to
\infty$. Therefore, any stable steady solution $F_i(t) \equiv F_i^{\ast}$
satisfies $\sum_{i=1}^{N} F_i^{\ast} = 0$.

\subsection {Single-level solution (egalitarian solution)}

It is easy to see that $F_i(t) \equiv 0 \,\,(i=1,\dots,N)$ is a steady solution
of Eq.~(\ref{e.mean-field-model.non-dimension}) for any values of $\rho \geq
0$. The Jacobian of the right side of
Eq.~(\ref{e.mean-field-model.non-dimension}) at this solution is given by a
circulant matrix
\begin{equation}
  \label{e.J1}
  J_{1,N} (a) =
  \begin{pmatrix}
    a      & b      & \cdots & b      & b      \\[-.2cm]
    b      & a      & \ddots & b      & b      \\[-.1cm]
    \vdots & \ddots & \ddots & \ddots & \vdots \\[-.1cm]
    b      & b      & \ddots & a      & b      \\[-.0cm]
    b      & b      & \cdots & b      & a
  \end{pmatrix},
\end{equation}
where $a:= \rho/2-1$ and $b:=-\rho/(2N')$. For later use, the Jacobian is
denoted as $J_{1,N} (a)$ to indicate that it is a (matrix-valued) function of
$a$. The eigenvalues of $J_{1,N} (a)$ are given by
\begin{align}
  \lambda
  \label{e.J1N.eigenvalue}
  &=
  a +  (N-1) b,\,\,
  a-b\\[0.1cm]
  \label{e.single-level.eigenvalue}
  &=
  -1, \,\,
  \frac{N}{2N'}\rho-1.
\end{align}
The multiplicity of the first eigenvalue $a+(N-1)b$ is $1$ and that of the
second $a-b$ is $N-1$.

A steady solution is linearly stable, if all the eigenvalues of the Jacobian
are negative \cite{strogatz94}. Thus, according to
Eq.~(\ref{e.single-level.eigenvalue}), the solution $F_i(t) \equiv 0$ is
linearly stable, if $\rho$ satisfies
\begin{equation}
  \label{e.rho_c}
  \rho < \frac {2N'}{N} =: \rho_c,
\end{equation}
where we define the critical value $\rho_c$. Note that, for the general case
with $F_0 \neq 1$ and $T_0 \neq 1$, this definition becomes
\begin{equation}
  \label{e.rho_c.general}
\rho_c := \frac {2N'}{N} \frac {F_0}{T_0}. 
\end{equation}
For $\rho > \rho_c$, the single-level solution is
unstable. This is consistent with the corresponding result in
Ref.~\cite{bonabeau95}. Note that $N-1$ eigenvectors associated with the second
eigenvalue $\lambda= \rho/\rho_c - 1$ become unstable simultaneously at
$\rho = \rho_c$.
%% It is obvious that $\lambda=-1$ corresponds to the dynamics of the total
%% fitness $S(t)$.

\subsection {Two-level solution}\label{s.two-level}
%subsubsection {existence}
%Hereafter, we assume that $N$ is an even number. Then,
Two-level asymmetric solutions have been studied in previous works
\cite{lacasa06, posfai18}, but in these studies, asymmetry is incorporated
directly into the model equations. Here, however, we show that there exist
stable asymmetric solutions even in the symmetric model given by
Eq.~(\ref{e.mean-field-model.non-dimension}) (Asymmetry is incorporated through
the initial condition). Moreover, linearly stable ranges in terms of $\rho$ are
derived for these asymmetric solutions.

Let us study steady two-level solutions with asymmetry: 
\begin{align}
  \label{e.mean-field.solution.two-state}
  F_i(t) \equiv
  \begin{cases}
    F^u  & (i \leq m), \\
    -F^l & (i > m),
  \end{cases}
\end{align}
where the constants $F^{u}, F^{l}$ are positive $F^{u}, F^{l} > 0$ and
$m = 1,\dots,N/2$. This parameter $m$ is the number of the upper-level agents
[$F_i(t) \equiv F^u$]; $m=N/2$ corresponds to a symmetric two-level solution.
For $m>N/2$, asymmetric solutions similar to those for $m<N/2$ exist, because of
the symmetry in Eq.~(\ref{e.mean-field-model.non-dimension}) with respect to
$F_i(t) \rightarrow -F_i(t)$. However, we omit these cases $m>N/2$ for brevity
of presentation.

The values of $F^{u}$ and $F^{l}$ can be determined by setting the right side of
Eq.~(\ref{e.mean-field-model.non-dimension}) zero. Thus, we obtain
\begin{align}
  \label{e.two-level.F1.F2}
  \begin{cases}
    \rho\frac {N-m}{N'} f(F^{u}+F^{l}) - F^{u}   & = 0, \\[0.1cm]
    \rho\frac {m}{N'}   f(F^{u} + F^{l}) - F^{l} & = 0.
  \end{cases}
\end{align}
If $F^u+F^l \leq 2$, then $f(F^u+F^l) = (F^u+F^l)/2$, and therefore we have
$F^u=F^l=0$ from Eq.~(\ref{e.two-level.F1.F2}) . Thus, $F^u+F^l > 2$ is
necessary for the existence of the two-level solutions. Under this condition,
Eq.~(\ref{e.two-level.F1.F2}) can be solved as
\begin{equation}
  \label{e.mean-field.solution.two-state.F}
  F^u = \rho\frac {N-m}{N'},\quad
  F^l = \rho\frac {m}{N'}.
\end{equation}
Since $F^u+F^l = 2\rho/\rho_c> 2$, the two-level solution exists for
$\rho > \rho_c$. 

%subsubsection {fig4}

\begin{figure}[t]
  \centerline{\includegraphics[width=8.0cm]{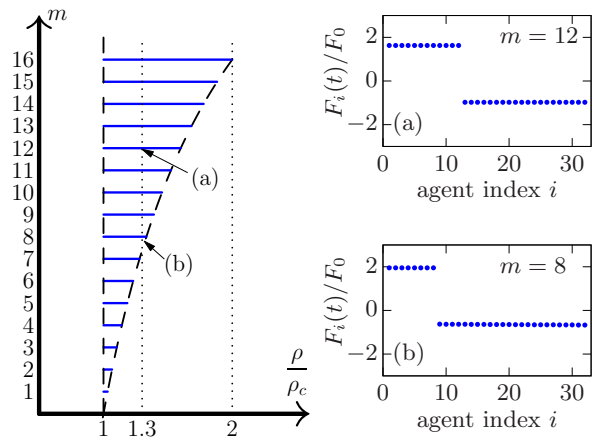}}
  %% \begin{minipage}[bt]{4.0cm}
  %%   \centerline{\includegraphics[width=3.97cm]{bullying.2level.eps}}
  %% \end{minipage}
  %% \begin{minipage}[bt]{3.5cm}
  %%   \centerline{\includegraphics[width=3.5cm]{m12.f.eps}}
  %%   \centerline{\includegraphics[width=3.5cm]{m8.f.eps}}
  %% \end{minipage}
  \caption{\label{f.2level.m-rho} (Left) Phase diagram ($\rho$ vs $m$) of
    two-level stable solutions. The total number of the agents is $N=32$. On the
    horizontal solid lines, the two-level solutions are stable. Dashed lines are
    theoretical prediction Eq.~(\ref{e.two-level-range}). Arrows indicate the
    parameter values used in the right figures. (Right) Examples of asymmetric
    two-level solutions obtained by numerical simulations. The density $\rho$ is
    set as $\rho/\rho_c=1.3$. The number of the upper-level agents $m$ is (a)
    $m=12$, and (b) $m=8$. A weakly hierarchical state $F_i(0)$, similar to the
    one shown in Fig.~\ref{f.agent.two-level}(a), is used as the initial
    condition, which is sorted as $F_i(0) > F_j(0)$ for $i < j$. Note that this
    order of $F_i(t)$ does not change with $t$ in the mean-field model [i.e.,
    $i_{\alpha}(t) \equiv \alpha$], thus the agent index $i$ is used as the
    horizontal axis.}
\end{figure}

%subsubsection {linear stability}
Linear stability analysis can be carried out in the same way as the previous
subsection. The Jacobian at the two-level steady solutions is given by
\begin{equation}
  \label{e.J2}
  J_{2,m}  =
  \begin{pmatrix}
    J_{1,m}(a) & O \\
    O   & J_{1,N-m}(a')
  \end{pmatrix},
\end{equation}
where $J_{1,m}(a)$ is an $m \times m$ matrix of the form of Eq.~(\ref{e.J1}) but
with $a=\rho(m-1)/(2N')-1$ [$b$ is the same as that in Eq.~(\ref{e.J1})],
$J_{1,N-m}(a')$ is an $(N-m) \times (N-m)$ matrix with
$a' = \rho (N-m-1)/(2N') - 1$, and $O$ is a zero matrix. By using
Eq.~(\ref{e.J1N.eigenvalue}), it is easy to find the eigenvalues of $J_{2,m}$ as
\begin{equation}
  \label{e.two-level.eigenvalues}
  \lambda = -1, \,\,
  \rho \frac {m}{2N'}-1,\,\,
  \rho \frac {N-m}{2N'}-1,
\end{equation}
with multiplicities $2, m-1$, and $N-m-1$, respectively. Therefore, the
two-level stable solution with $m$ exists for $\rho$ satisfying
\begin{equation}
  \label{e.two-level-range}
  1 < \frac {\rho}{\rho_c} < \frac {N}{N-m}.
\end{equation}
Thus, at $\rho = \rho_c$, the steady solution of
Eq.~(\ref{e.mean-field-model.non-dimension}) changes abruptly from
$F_i(t) \equiv 0$ to the above values in
Eq.~(\ref{e.mean-field.solution.two-state.F}). This discontinuity originates
from the fact that $f(x)$ is not differentiable.  Even for
$\rho/\rho_c > N/(N-m) $, the two-level solutions with $m$ exist, but they are
unstable because the third eigenvalue in Eq.~(\ref{e.two-level.eigenvalues})
becomes positive.

%subsubsection {explanation of figure}

In Fig.~\ref{f.2level.m-rho}, the ranges of $\rho$ where a stable two-level
solution exists are displayed by horizontal lines. The symmetric solution
($m=N/2$) has the widest stable range; the stable range is shorter for stronger
asymmetry (i.e., for smaller values of $m$). Asymmetric solutions shown in
Figs.~\ref{f.2level.m-rho}~(a) and (b) are obtained by numerical simulations;
these solutions resemble the result for the agent model shown in
Fig.~\ref{f.agent.two-level}(b).

As shown in Fig.~\ref{f.2level.m-rho}(Left), the two-level solutions
($m=1,\dots,N-1$) appear simultaneously at $\rho = \rho_c$ through bifurcations
of the pitchfork type (though there is a discontinuity).  This can be easily
checked by setting $F_i(t) = F^{u}(t)$ for ($i\leq m$), and $F_i(t) = -F^{l}(t)$
for ($i>m$); this form of the trajectory $F_i(t)$ is a solution in an invariant
two-dimensional subspace. If we define $\Delta F(t)= F^{u}(t) + F^{l}(t)$, it is
easy to show that
\begin{equation}
  \label{e.invariant_subspace.two-level}
  \frac {d \Delta F (t)}{dt} = -\Delta F(t)
  +
  \frac {2\rho}{\rho_c} f\bigl(\Delta F(t)\bigr).
\end{equation}
Examining the functional form of the right side [as a function of $\Delta F(t)$]
for $\rho < \rho_c$ and $\rho > \rho_c$, it is found that the bifurcation at
$\rho=\rho_c$ is the superciritical pitchfork type \cite{strogatz94}. Note
however that this analysis in invariant subspaces is insufficient for a proof of
the linear stability. See Appendix \ref{s.unstable-three-level} for a similar
argument on three-level solutions, for which two stable solutions appear also
through the superciritical pitchfork bifurcations, but they are unstable in some
directions perpendicular to the invariant subspaces.

\subsection {Three-level solution}\label{s.three-level}
%paragraph {fig5}

\begin{figure}[t]
  \centerline{\includegraphics[width=8.5cm]{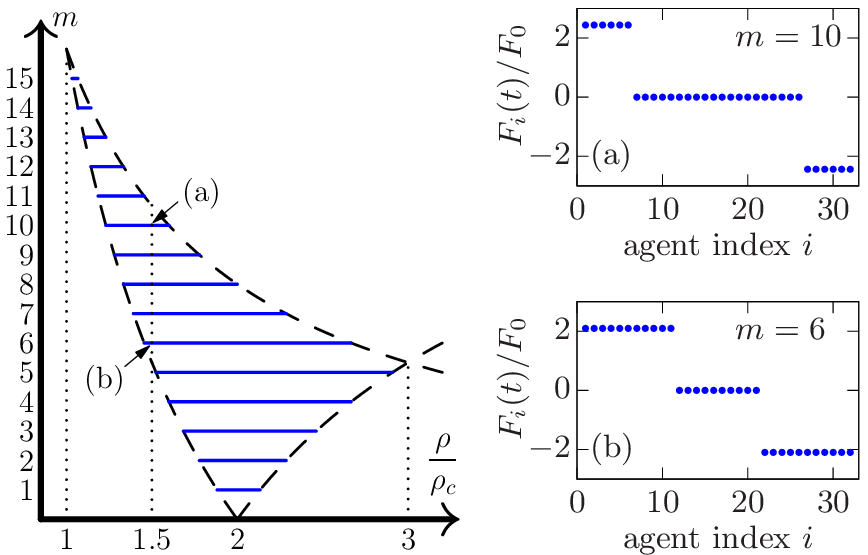}}
  %% \begin{minipage}[bt]{4.8cm}
  %%   \centerline{\includegraphics[width=4.8cm]{bullying.3level.eps}}
  %% \end{minipage}
  %% \hspace*{.1cm}
  %% \begin{minipage}[bt]{3.5cm}
  %%   \centerline{\includegraphics[width=3.5cm]{m10.f.eps}}
  %%   \centerline{\includegraphics[width=3.5cm]{m6.f.eps}}
  %% \end{minipage}
  \caption{\label{f.3level.m-rho} (Left) Phase diagram ($\rho$ vs $m$) of
    three-level stable solutions. The total number of the agents is $N=32$. On
    the horizontal solid lines, the three-level solutions are stable. Dashed
    lines are theoretical prediction Eq.~(\ref{e.three-state.range}). Arrows
    indicate the parameter values used in the right figures. (Right) Examples of
    three-level solutions obtained by numerical simulations. The density $\rho$
    is set as $\rho/\rho_c=1.5$. Half the number of the middle-level agents $m$
    is (a) $m=10$, and (b) $m=6$. Weakly hierarchical states, similar to the one
    shown in Fig.~\ref{f.agent.three-level}(a), are used as initial conditions.}
\end{figure}

%paragraph {theory}

There are also many three-level steady solutions, and thus here we focus only on
symmetric three-level solutions.  In this subsection, steady solutions of the
following form are shown to be stable:
\begin{align}
  \label{e.mean-field.solution.three-state}
  F_i(t) \equiv
  \begin{cases}
    F  & \left(1 \leq i \leq \frac {N}{2} - m\right),                    \\[.1cm]
    0  & \left(\frac {N}{2} - m < i \leq \frac {N}{2} + m\right), \\[.1cm]
    -F & \left(\frac {N}{2} + m < i \leq N \right).
  \end{cases}
\end{align}
Here, $2m$ is the number of the middle-level agents, for which $F_i(t)\equiv 0$;
therefore $m$ should satisfy $0 < m < N/2$. Moreover, the constant $F$ is
assumed to satisfy $F>2$ (even if $F<2$, there exist some steady solutions,
but they are linearly unstable. See Appendix \ref{s.unstable-three-level}).
Substituting Eq.~(\ref{e.mean-field.solution.three-state}) into the right side
of Eq.~(\ref{e.mean-field-model.non-dimension}), we found that
\begin{equation}
  \label{e.mean-field.solution.three-state.F}
  F = \frac {\rho}{\rho_c} + \frac {\rho m}{N'}.
\end{equation}
Since we assume $F>2$, the steady solution
[Eq.~(\ref{e.mean-field.solution.three-state.F})] exists for
$\rho > 4N'/(N+2m)$.

The Jacobian of these steady solutions is given by
\begin{equation}
  \label{e.J3m}
  J_{3,m}^{F>2}  =
  \begin{pmatrix}
    J_{1,N/2-m}(a) & O           & O \\[.2cm]
    O              & J_{1,2m}(a') & O \\[.2cm]
    O              & O           & J_{1,N/2-m}(a)
  \end{pmatrix},
\end{equation}
where $J_{1,N/2-m}(a)$ is an $(N/2-m) \times (N/2-m)$ matrix of the form of
Eq.~(\ref{e.J1}) with $a=\rho(N'-1-2m)/(4N')-1$, and $J_{1,m}(a')$ is an
$2m \times 2m$ matrix with $a=\rho(2m-1)/(2N')-1$. By using
Eq.~(\ref{e.J1N.eigenvalue}), we obtain the eigenvalues of $J_{3,m}^{F>2}$ as 
\begin{equation}
  \label{e.three-level.eigenvalues}
  \lambda =
  -1,\,\,
  \rho \frac {N-2m}{4N'} -1,\,\,
  \rho \frac {m}{N'} -1,
\end{equation}
with multiplicities $3, N-2m-2$, and $2m-1$, respectively. Therefore, the
three-level stable solution with $m$
[Eq.~(\ref{e.mean-field.solution.three-state})] exists for $\rho$ satisfying
\begin{equation}
  \label{e.three-state.range}
  \frac {2N}{N+2m}
  <
  \frac {\rho}{\rho_c}
  <
  \mathrm{max}
  \left(\frac {2N}{N-2m}, \,\,\frac {N}{2m}\right).
\end{equation}
Even for $\rho/\rho_c$ larger than this upper bound, the three-level solutions
exist, but they are unstable because the second or the third eigenvalues in
Eq.~(\ref{e.three-level.eigenvalues}) become positive.

%paragraph {explanation of fig}

In Fig.~\ref{f.3level.m-rho}, the ranges of $\rho$ where the stable three-level
solutions exist are displayed by horizontal lines. The widest stable range is at
$m=N/6$, at which the three levels have the equal numbers of agents (the example
shown in the figure is for $N=32$, and thus $N/6$ is not an integer. If $N$ is a
multiple of $3$, there is a steady solution for which each level has $N/3$
agents). In Figs.~\ref{f.3level.m-rho} (a) and (b), solutions obtained by
numerical simulations are displayed. These solutions resemble the result for the
agent model shown in Fig.~\ref{f.agent.three-level}(b).

\subsection {$N$-level solution (linear hierarchy)} \label{s.n-level}

Linear hierarchies are frequently observed in animal societies. In a linear
hierarchy, if an individual A dominates B and B dominates C, then A dominates C
\cite{chase02} (i.e., a transitive relationship). At high values of $\rho$,
there exists a steady $N$-level solution, in which each agent has a different
values of $F_i(t)$. This completely stratified solution is reminiscent of the
linear hierarchy.

Here, let us assume the following solution
\begin{equation}
  \label{e.solution.N-level}
  F_i (t) \equiv F - \frac {2F}{N'}(i-1), \quad (i=1,\dots,N),
\end{equation}
where $F$ is a constant to be determined, and we also assume $F > N'$. In order
that the above $F_i(t)$ is a steady solution, i.e., $dF_i(t)/dt \equiv 0$ in
Eq.~(\ref{e.mean-field-model.non-dimension}), $F$ should satisfy
\begin{equation}
  \label{e.complete.stratify}
  F = \rho.
\end{equation}
In the derivation, we used the assumption $F > N'$ as
$F_i(t) - F_j(t) = 2F(j-i)/N' > 2F/N' > 2$, where $i<j$. From $F > N'$ and
Eq.~(\ref{e.complete.stratify}), $\rho$ should also satisfy $\rho>N'$, or
\begin{equation}
  \label{e.stable-range-N-level}
  \frac {\rho}{\rho_c} > \frac {N}{2}.
\end{equation}
Thus, the $N$-level solution exists only at large $\rho$.

The stability of the $N$-level solution is easy to prove. The Jacobian of this
steady solution is simply given by $J_N = -I$, where $I$ is the $N \times N$
identity matrix. Therefore, the $N$-level solution
[Eq.~(\ref{e.solution.N-level})] is stable.

\section {Ergodicity in agent model}\label{s.ergodicity-breaking}
%subsection {definitions}
As shown in Figs.~\ref{f.agent.two-level} and \ref{f.agent.three-level}, the
agent model behaves similarly to the mean-field model for $1/\gamma \ll T_0$ and
$\eta \ll F_0$. But, if these conditions are not fulfilled, the agent model behaves
differently from the mean-field model. In this section, the dependence of the
agent model on these parameters $\gamma$ and $\eta$ is numerically studied.

As a quantity characterizing the dynamics of the agent model, we use the
standard deviation $\sigma(\gamma, \eta)$ of the time-averaged DS,
$\overline{F}_i$, defined as
\begin{align}
  \label{e.mu(rho,eta)}
  \mu(\gamma, \eta)
  &:=
  \frac {1}{N} \sum_{i=1}^{N} \overline{F}_i,\\
  \label{e.sigma(rho,eta)}
  \sigma^2(\gamma, \eta)
  &:=
  \frac {1}{N} \sum_{i=1}^{N} [\overline{F}_i - \mu(\gamma, \eta)]^2.  
\end{align}
The time averaged DS, $\overline{F}_i$, is defined as
\begin{equation}
  \label{e.overline(F)_i}
  \overline{F}_i := \frac {1}{T}\int_{0}^{T} F_i(t) dt,
\end{equation}
where the dynamics of $F_i(t)$ is given by Eq.~(\ref{e.agent_model.Fi}) with the
parameters $\eta$ and $\gamma$. Thus, the standard deviation
$\sigma(\gamma, \eta)$ also depends on these parameters.

\begin{figure}[t]
  \centerline{\includegraphics[width=8.8cm]{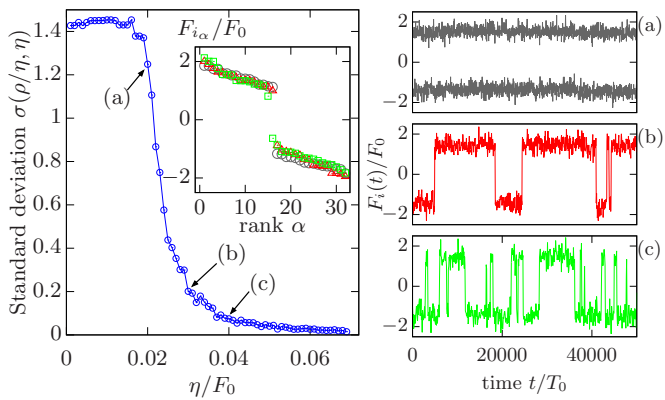}}
  %% \begin{minipage}[bt]{4.7cm}
  %%   \centerline{\includegraphics[width=4.7cm]{ergodicity.f.eps}}
  %% \end{minipage}
  %% \hspace*{.2cm}
  %% \begin{minipage}[bt]{3.5cm}
  %%   \vspace*{-.25cm}
  %%   \centerline{\includegraphics[width=3.5cm]{eta002.f.eps}}
  %%   \vspace*{.13cm}
  %%   \centerline{\includegraphics[width=3.5cm]{eta003.f.eps}}
  %%   \vspace*{.13cm}
  %%   \centerline{\includegraphics[width=3.5cm]{eta004.f.eps}}
  %% \end{minipage}
  \caption{\label{f.ergodicity} (Left) Standard deviation
    $\sigma(\rho/\eta, \eta)$ vs $\eta$ in the agent model with $N = 32$. The
    value of $\rho$ is fixed as $\rho=1.5\rho_c$, for which the egalitarian
    solution $F_i(t) \equiv 0$ is unstable in the mean-field
    approximation. Arrows indicate the parameter values used in the right
    figures. The time average in Eq.~(\ref{e.overline(F)_i}) is taken over a
    time interval during which $10^9$ contests occur. The inset is a snapshot of
    $F_{i_{\alpha}}(t)$ vs rank $\alpha$ for $\eta = 0.02F_0$ (circle),
    $\eta = 0.03F_0$ (triangle), and $\eta = 0.04F_0$ (square). (Right) Typical
    trajectories of $F_i(t)$ for (a) $\eta = 0.02F_0$, (b) $\eta = 0.03F_0$, and
    (c) $\eta = 0.04F_0$. In (a), two trajectories are displayed, whereas a
    single trajectory is displayed in (b) and (c). }
\end{figure}

If the system is ergodic, a time average tends to a single value, which is equal
to the ensemble average, in a long time limit ($T \to \infty$). In the agent
model [Eqs.~(\ref{e.agent_model.Fi}) and (\ref{e.agent_model.Fj})], all the
agents are equivalent, and therefore the limiting value is the same for all the
agents, and thus it follows that $\sigma(\gamma, \eta)$ vanishes at
$T \to \infty$.  Accordingly, $\sigma(\gamma, \eta)$ can be used as a parameter
of ergodicity breaking \cite{he08, miyaguchi13}. In Fig.~\ref{f.ergodicity}, we
set $\gamma \eta$ is constant (i.e., $\rho=\gamma \eta$ is constant) to fix the
corresponding mean-field model [See Eq.~(\ref{e.def.mean-field-model})], and
numerically obtain the variance $\sigma^2(\rho/\eta, \eta)$ as a function of
$\eta$.

%subsection {fig}

%subsection {explanation of fig}
As shown in Fig.~\ref{f.ergodicity} (Left), the standard deviation
$\sigma(\rho/\eta, \eta)$ is far away from zero for small values of $\eta$. In
fact, the agents are separated into two groups as shown in the inset of
Fig.~\ref{f.ergodicity}(Left); these two groups correspond to the two-level
solution in the mean-field model with $m=N/2$
[Eq.~(\ref{e.mean-field.solution.two-state})]. For small $\eta$, the members of
these two groups rarely change in the course of time evolution, as shown in
Fig.~\ref{f.ergodicity}(a), where two typical trajectories $F_i(t)$ are
displayed.

For large values of $\eta$, the agents are still separated into two groups again
[See the inset of Fig.~\ref{f.ergodicity}(Left)], but the agents frequently move
from one group to the other as shown in Figs.~\ref{f.ergodicity}(b) and
(c). Accordingly, all the time averages $\overline{F}_i$ ($i=1,\dots,N$) tend to
zero as $T$ increases, and therefore the standard deviation
$\sigma(\rho/\eta, \eta)$ also vanishes as shown in Fig.~\ref{f.ergodicity}
(Left). The transitions of the agents from one group to the other occur, because
the impact of each contest becomes significant for large $\eta$ [though a time
average of this effect, given by $\gamma \eta$, is the same in all the numerical
simulations in Figs.~\ref{f.ergodicity} (a)--(c)]. It should be also noted that,
even though the time average $\overline{F}_i$ vanishes at large $\eta$, a
hierarchy exists in snapshots $F_i(t)$ as shown in the inset of
Fig.~\ref{f.ergodicity}(Left), where the agents are separated into two groups,
and thus the system is not egalitarian.

At small $\eta$, the ergodicity seems to be violated as shown in
Fig.~\ref{f.ergodicity}(a). However, it is probable that it just takes too long
time to observe transitions of the agents from one group to the other, and thus
the ergodicity might not be violated. This is because a sequence of contests at
large $\eta$ which causes a transition of an agent can be possible, in
principle, to occur even at small $\eta$ (though the probability of occurrence
of such sequence of contests is quite small). Therefore, the observed violation
of the ergodicity might well be just apparent.

\section {Discussion}\label{s.summary}

Since the appearance of the seminal paper \cite{bonabeau95}, the Bonabeau model
has been employed to explain experimental data of animal hierarchy formations,
and many modified versions have been proposed \cite{stauffer03a, stauffer03b,
  odagaki06, okubo07, tsujiguchi07}. But, understanding of the original Bonabeau
model has not been far from satisfactory due to difficulty in treating its
nonlinearity. In this paper, a piecewise linear version of the Bonabeau model
was introduced. By using the mean-field approximation, it was shown that there
are many asymmetric solutions, and that coexistence of the stable solutions
takes place. In addition, an apparent transition in ergodic behaviors is found
in the agent model.

Our model assumed that encounters of the agents are completely random. Namely,
at each contest time $t_n$, the agents $i$ and $j$ are randomly chosen from the
$N$ agents. But, it is known that if the agents $i$ and $j$ contest, then these
agents $i$ and $j$ are more likely to contest in the next contest event than
other agents \cite{lindquist09}. Remarkably, it is also shown in
Ref.~\cite{lindquist09} that the persistent time during which the same
individuals successively contest follows a power-law distribution. Such a
non-Markovian memory effect can be easily implemented in the agent model, by
introducing a persistent-time distribution \cite{miyaguchi13, miyaguchi19}
\begin{equation}
  \label{e.persistent-time.pdf}
  w_{\mathrm{p}}(\tilde{\tau}) \simeq \frac {a}{\tilde{\tau}^{1+\alpha}}
  \quad (\tau \to \infty),
\end{equation}
where $a$ and $\alpha$ are positive constants. We choose a sequences of
persistent times $\tilde{\tau}_1, \tilde{\tau}_2, \cdots$, each following
$w_{\mathrm{p}}(\tilde{\tau})$, and define renewal times as
$\tilde{t}_n := \sum_{k=1}^{n} \tilde{\tau}_{k}$, at which the contestants
change. In each interval $[\tilde{t}_{n-1}, \tilde{t}_n]$, the same agents $i$
and $j$ contest. This generalized model should be studied in future works.

The linear hierarchy, frequently observed in animal societies, is characterized
by the transitive relationship (See Sec.~\ref{s.n-level}); however, intransitive
relationships are also observed by suppressing group processes
\cite{chase02}. Such intransitive relationships cannot be described by the
Bonabeau model, because it is always transitive from its definition; i.e., if
$F_i(t) > F_j(t)$ and $F_j(t) > F_k(t)$, then $F_i(t) > F_k(t)$. To describe the
intransitive relationships, it is necessary to introduce an anti-symmetric
matrix $F_{ij}(t)$ which describes the dominance relationship between $i$ and
$j$. In the Bonabeau model, $F_{ij}(t)$ could be defined by
$F_{ij}(t):= F_i(t) - F_j(t)$, but the matrix $F_{ij}(t)$ cannot be described by
a single vector in general.

Therefore, future work is needed to develop a generalized model for $F_{ij}(t)$,
and to elucidate how the transitive relationship [i.e., if $F_{ij}(t) > 0$ and
$F_{jk}(t) > 0$, then $F_{ik}(t) > 0$] emerges (or self-organizes). In such a
generalized model, a bystander effect should be incorporated, in addition to the
winner/loser effects \cite{chase02, grosenick07}. The bystander effect is a
mechanism that an individual who witnesses a contest between other individuals
is influenced by the result of that contest; the witness might learn its status
vicariously by observing contests between other individuals \cite{grosenick07}.
Without such a bystander effect, intransitive relationships should be frequently
observed \cite{chase02}.

Finally, we neglect the noise terms in Eqs.~(\ref{e.agent_model.Fi}) and
(\ref{e.agent_model.Fj}) in this paper, and thus contest dynamics is purely
deterministic except the random choice of two contestants. In real societies,
however, contestants have some random factors such as their physical conditions.
Therefore, the noise terms might be important and should be studied in future
works.

%section {acknowledgments}

%% \begin{acknowledgments}
%%   This work was supported by Grant-in-Aid (KAKENHI) for Scientific Research C
%%   (Grand No. JP18K03417).
%% \end{acknowledgments}

\appendix { }
\section {Rescaling}\label{s.rescale}

In this Appendix, the agent model and the mean-field-model are transformed into
simpler forms by introducing rescaled variables.  Let us define the rescaled
(non-dimensional) variables as
\begin{align}
  \label{e.non_dimensional_vars}
  \bar{t} = \frac {t}{T_0}, \quad
  \bar{F}_i(\bar{t}) = \frac {F_i (t)}{F_0}.
\end{align}
Then, Eqs.~(\ref{e.agent_model.Fi}) and (\ref{e.agent_model.Fj}) can be rewritten
as (we omit the noise terms)
\begin{align}
  \label{e.agent_model.Fi.nondimension}
  \bar{F}_i(\bar{t}_n^+) &= \bar{F}_i(\bar{t}_n^-)
  + \bar{\eta} \bar{f}(\bar{F}_i(\bar{t}^-_n) - \bar{F}_j(\bar{t}^-_n)),
  \\[0.1cm]
  \bar{F}_j(\bar{t}_n^+) &= \bar{F}_j(\bar{t}_n^-)
  + \bar{\eta} \bar{f}(\bar{F}_j(\bar{t}^-_n) - \bar{F}_i(\bar{t}^-_n)),
\end{align}
where $\bar{\eta}$ and $\bar{f}(x)$ are defined respectively as
$\bar{\eta} = \eta/F_0$ and
\begin{align}
  \label{e.def.f(x).nondimension}
\bar{f}(x)=
\begin{cases}
  -1          & (x<-2),
 \\[0.1cm]
  \frac{x}{2} & (-2\leq x \leq 2),
 \\[0.1cm]
  1           & (x > 2).
\end{cases}
\end{align}

The exponential distribution of the intervals $\tau$ [Eq.~(\ref{e.exp.dist})] is
also rescaled as
\begin{equation}
  \label{e.exp.dist.nondimension}
  \bar{w}(\bar{\tau}) =  \bar{\gamma}_a e^{- \bar{\gamma}_a \bar{\tau}},
\end{equation}
where $\bar{\gamma}_a = \gamma_a T_0$. The relaxation dynamics
[Eq.~(\ref{e.relaxation})] is simply given by
\begin{equation}
  \label{e.relaxation.nondimension}
  \frac {d\bar{F}_i(\bar{t})}{d\bar{t}} = - \bar{F}_i(\bar{t}).
\end{equation}
Therefore, the two parameters $\bar{\eta}$ and $\bar{\gamma}_a$ completely
characterize the agent model.

Similarly, by using the transformations in Eq.~(\ref{e.non_dimensional_vars}),
the mean-field model in Eq.~(\ref{e.def.mean-field-model}) becomes
\begin{align}
  \label{e.mean-field-model.non-dimension.app}
  \frac {d\bar{F}_i(\bar{t})}{d\bar{t}}
  =
  \frac {\bar{\rho}}{N'}
  \sum_{\begin{subarray}{c}j=1\\j\neq i\end{subarray}}^{N}
  \bar{f} \bigl(\bar{F}_i(\bar{t}) - \bar{F}_j(\bar{t})\bigr)
  -
  \bar{F}_{i}(\bar{t}).
\end{align}
where $\bar{\rho}$ is defined as $\bar{\rho}= \rho T_0/F_0$. Thus, $\bar{\rho}$
is the only parameter of the mean-field model. Note that, even if $\bar{\rho}$
is constant, corresponding parameter values in the agent model ($\bar{\eta}$ and
$\bar{\gamma}_a$) are not uniquely determined, because
$\bar{\rho} = \bar{\gamma}\bar{\eta}$ with $\bar{\gamma} := 2\bar{\gamma}_a/N$.

\section {Unstable three-level solution}\label{s.unstable-three-level}
%subsection {existence}

In Sec.~\ref{s.three-level}, we study stable three-level solutions
[Eq.~(\ref{e.mean-field.solution.three-state})], but there also exist unstable
three-level solutions. In this Appendix, we show that the three-level unstable
solutions emerge simultaneously at $\rho=\rho_c$, and these unstable solutions
become stable at some values of $\rho > \rho_c$.

First, it is easy to show that a steady three-level solution of the form of
Eq.~(\ref{e.mean-field.solution.three-state}) does not exist for $0 < F < 1$,
and we already study the three-level solutions for $F>2$ in
Sec.~\ref{s.three-level}. Thus, here we assume $1<F<2$.
For $1<F<2$, the three-level solution is given by
\begin{equation}
  \label{e.three-level.solution.1<F<2}
  F= \frac {\rho}{2} \frac {N-2m}{N' - \rho m},
\end{equation}
with $m=0,\dots,N/2-1$ ($m=0$ corresponds to the symmetric two-level
solution). Accordingly, the range of $\rho$ satisfying $1<F<2$ is given by
\begin{equation}
  \label{e.range-exist.1<F<2}
  1 < \frac {\rho}{\rho_c} < \frac {2N}{N+2m}.
\end{equation}
Note that the upper bound is equivalent to the lower bound of the stable
three-level solution [Eq.~(\ref{e.three-state.range})]. In fact, the stability
of the three-level solutions for each value of $m$ changes at
$\rho/\rho_c = 2N/(N+2m)$ as shown below. The bifurcation at this point might be
subcritical-pitchfork type (however, we should note that details of this
bifurcation remain still unclear).

%subsection {fig}

\begin{figure}[t!]
  \centerline{\includegraphics[width=7.2cm]{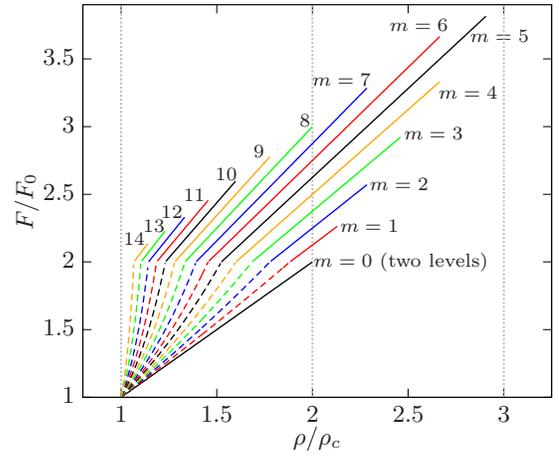}}
  %\centerline{\includegraphics[width=7.2cm]{bullying-3level-phase-dia.eps}}
  \caption{\label{f.three-level.phase-dia}Dominance score $F$ for two-level and
    three-level symmetric solutions as functions of $\rho$. The number of agents
    are set as $N=32$. The numbers in the figure are the corresponding values of
    $m$. The dashed curves are unstable solutions given by
    Eq.~(\ref{e.three-level.solution.1<F<2}), and the solid curves are stable
    two-level and three level solutions given by
    Eq.~(\ref{e.mean-field.solution.two-state.F}) with $m=N/2$ and
    Eq.~(\ref{e.mean-field.solution.three-state.F}), respectively. Beyond these
    stable ranges, all the solutions with $m$ still exist, but they are unstable
    (See Secs.~\ref{s.two-level} and \ref{s.three-level}). These unstable ranges
    are not shown for brevity.}
\end{figure}

%subsection {stability}

Next, the stability of the three-level solutions
[Eq.~(\ref{e.three-level.solution.1<F<2})] is elucidated. In this case, the
Jacobian matrix is given by
\begin{equation}
  \label{e.J3m.1<F<2}
  J_{3,m}^{1<F<2}  =
  \begin{pmatrix}
    J_{1,N/2-m}(a) & B            & O   \\[.2cm]
    B^t            & J_{1,2m}(a') & B^t \\[.2cm]
    O              & B            & J_{1,N/2-m}(a)
  \end{pmatrix},
\end{equation}
where $J_{1,N/2-m}$ is an $(N/2-m)\times (N/2-m)$ matrix of the form of
Eq.~(\ref{e.J1}) with $a=\rho (N'+2m-1)/(4N')-1$, $J_{1,2m}(a')$ is a
$2m\times 2m$ matrix with $a'=\rho/2-1$, and $B$ is an $(N/2-m)\times 2m$
matrix, all the elements of which are the same and given by $-\rho/(2N')$. $B^t$
is the transpose of $B$.

After a somewhat lengthy but elementary calculation, we obtain four eigenvalues
of the Jacobian in Eq.~(\ref{e.J3m.1<F<2}). Two of the four eigenvalues are
given by
\begin{align}
  \label{e.eigenvalue.1<F<2.1}
  \lambda
  =
  \frac {\rho}{\rho_c} - 1, \quad
  \rho\frac {N+2m}{4N'} - 1,
\end{align}
with multiplicities $2m$ and $N-2(m+1)$, respectively. The first eigenvalue in
Eq.~(\ref{e.eigenvalue.1<F<2.1}) is positive because of
Eq.~(\ref{e.range-exist.1<F<2}). Therefore, the three-level solutions in
Eq.~(\ref{e.three-level.solution.1<F<2}) are unstable. Note however that the
first eigenvalue does not exist for $m=0$ (i.e., for the two-level symmetric
solution), because the multiplicity becomes zero, and therefore it is not
contradicting with the fact that the two-level solution is stable (See
Sec.~\ref{s.two-level}). Note also that the second eigenvalue is negative, and
it does not exist, if $m=N/2-1$. The remaining two eigenvalues are given by
\begin{equation}
  \label{e.eigenvalue.1<F<2.2}
  \lambda =
  - 1,\quad
  \rho \frac {m}{N'} - 1.
\end{equation}
These eigenvalues are simple and negative.

%subsection {explanation of fig}

A phase diagram of the symmetric two- and three-level solutions are displayed in
Fig.~\ref{f.three-level.phase-dia}. These steady solutions emerge at
$\rho = \rho_c$, but only two-level solution is stable, and all the three-level
solutions are unstable. For $F>2$, however, the two-level solution becomes
unstable (See Sec.~\ref{s.two-level}), whereas the three-level solutions, which
are given by Eq.~(\ref{e.mean-field.solution.three-state.F}), become stable.

%subsection {fig}

\begin{figure}[t!]
  \centerline{\includegraphics[width=7.0cm]{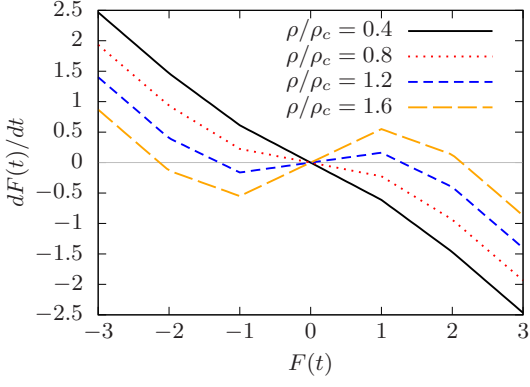}}
  %\centerline{\includegraphics[width=7.0cm]{bullying-3level-inv-space.eps}}
  \caption{\label{f.inv-space} $dF(t)/dt$ in
    Eq.~(\ref{e.invariant-subspace.each}) vs $F(t)$ for four different values of
    $\rho$: $\rho/\rho_c = 0.4$ (solid line), $0.8$ (dotted line), $1.2$ (dashed
    line), and $1.6$ (long-dashed line). $N$ and $m$ are set as $N=32$ and
    $m=6$. For $\rho < \rho_c$, $dF(t)/dt$ is monotonically decreasing, and thus
    the origin $F(t) \equiv 0$ is the stable fixed point. For $\rho > \rho_c$,
    however, the origin is unstable, and two stable fixed points, that
    correspond to Eq.~(\ref{e.three-level.solution.1<F<2}), appear. Note that
    these stable fixed points are stable only in the invariant subspace, and
    unstable in some directions perpendicular to this subspace.}
\end{figure}

%subsection {bifurcation}

Finally, let us consider how these solutions emerge. To elucidate this, we study
one-dimensional invariant subspaces described by the following solution
\begin{align}
  \label{e.F(t).invariant-subspace}
  F_i(t) =
  \begin{cases}
    F(t)  & \left(1 \leq i \leq \frac {N}{2} - m\right),             \\[.1cm]
    0     & \left(\frac {N}{2} - m < i \leq \frac {N}{2} + m\right), \\[.1cm]
    -F(t) & \left(\frac {N}{2} + m < i \leq N \right),
  \end{cases}
\end{align}
where $m=0,1,\dots,N/2-1$, and $F(t)$ can be either positive or negative. The
time evolution equation for $F(t)$ is obtained by inserting
Eq.~(\ref{e.F(t).invariant-subspace}) into
Eq.~(\ref{e.mean-field-model.non-dimension}) as
\begin{align}
  %\label{e.invariant-subspace}
  \frac {dF(t)}{dt}
  &=
  -F(t) + \frac {\rho}{N'}
  \left[
  2m f \bigl(F(t)\bigr) +
  \frac {N-2m}{2} f\bigl(2F(t)\bigr)
  \right]
  \notag\\[0.1cm]
  \label{e.invariant-subspace.each}
  &=
  \begin{cases}
    \left(\frac {\rho}{\rho_c} - 1\right) F(t)
    & [0<F(t)<1], \\[.2cm]
    \left(\rho\frac {m}{N'} - 1\right) F(t)  + \rho \frac {N-2m}{2N'}
    & [1<F(t)<2], \\[0.2cm]
    -F(t) + \rho \frac {N+2m}{2N'}
    & [2<F(t)],
  \end{cases}
\end{align}
where the equation only for $F(t)>0$ is explicitly given; the explicit
expression for $F(t)<0$ is readily obtained from the fact that $f(x)$ given in
Eq.~(\ref{e.def.f(x).nondimension}) is an odd function. Note also that the slope
$\rho m/N' - 1$ in Eq.~(\ref{e.invariant-subspace.each}), which is negative for
$\rho > \rho_c$, corresponds to the second eigenvalue in
Eq.~(\ref{e.eigenvalue.1<F<2.2}).

From the first equation in the right side of
Eq.~(\ref{e.invariant-subspace.each}), the single-level solution $F(t) \equiv 0$
is stable for $\rho < \rho_c$ and unstable for $\rho > \rho_c$. The two-level
($m=0$) and three-level ($m>0$) solutions emerge at $\rho=\rho_c$
simultaneously, and they are stable because of $\rho m/N' - 1 < 0$ for
$\rho > \rho_c$. Due to the symmetry, $-F(t)$ is also a solution in the
invariant subspaces, and thus there are two stable fixed points in each
invariant subspace with $m$.

%subsection {explanation fig}

This bifurcation is readily understood by a phase diagram \cite{strogatz94}
shown in Fig.~\ref{f.inv-space}, in which $dF(t)/dt$ in
Eq.~(\ref{e.invariant-subspace.each}) is displayed as a function of $F(t)$. It
is clear that the bifurcation at $\rho = \rho_c$ can be considered as a
superciritical pitchfork type. Although the bifurcations are pitchfork type and
thus the two emerged fixed points are stable, these fixed points except the
two-level solutions ($m=0$) are stable only in the invariant subspaces; in fact,
they are unstable in some directions perpendicular to the subspaces, because the
first eigenvalue in Eq.~(\ref{e.eigenvalue.1<F<2.1}) is positive.

%section {bibliography}

%\bibliography{paper}


\end {document}